\documentclass{article}

\usepackage{arxiv}

\usepackage[utf8]{inputenc} 
\usepackage[T1]{fontenc}    
\usepackage{hyperref}       
\usepackage{url}            
\usepackage{booktabs}       
\usepackage{amsfonts}       
\usepackage{nicefrac}       
\usepackage{microtype}      
\usepackage[dvipsnames]{xcolor}
\usepackage{graphicx}
\usepackage{siunitx}
\usepackage{multirow}
\usepackage{lipsum}

\usepackage{threeparttable}

\hypersetup{
    colorlinks = true,
}

\def\onedot{.}
\def\eg{\emph{e.g}\onedot}
\def\ie{\emph{i.e}\onedot}
\def\etal{\emph{et al}\onedot}

\title{\textbf{Learning domain-agnostic visual representation for computational pathology using medically-irrelevant style transfer augmentation}}

\author{Rikiya Yamashita, Jin Long, Snikitha Banda, Jeanne Shen, Daniel L. Rubin \\
Stanford University \\
\texttt{\{rikiya, jinlong, jeannes, rubin\}@stanford.edu}
}

\begin{document}
\maketitle

\begin{abstract}
Suboptimal generalization of machine learning models on unseen data is a key challenge which hampers the clinical applicability of such models to medical imaging. Although various methods such as domain adaptation and domain generalization have evolved to combat this challenge, learning robust and generalizable representations is core to medical image understanding, and continues to be a problem. Here, we propose \textbf{STRAP} (\textbf{S}tyle \textbf{TR}ansfer \textbf{A}ugmentation for histo\textbf{P}athology), a form of data augmentation based on random style transfer from non-medical style source such as artistic paintings, for learning domain-agnostic visual representations in computational pathology. Style transfer replaces the low-level texture content of an image with the uninformative style of randomly selected style source image, while preserving the original high-level semantic content. This improves robustness to domain shift and can be used as a simple yet powerful tool for learning domain-agnostic representations. We demonstrate that STRAP leads to state-of-the-art performance, particularly in the presence of domain shifts, on two particular classification tasks in computational pathology.
\end{abstract}

\newcommand\blfootnote[1]{
  \begin{NoHyper}
  \renewcommand\thefootnote{}\footnote{#1}%
  \addtocounter{footnote}{-1}%
  \end{NoHyper}
}
\blfootnote{Correspondence to \texttt{rikiya@stanford.edu}}
\blfootnote{Code: \textcolor{magenta}{\url{https://github.com/rikiyay/style-transfer-for-digital-pathology}}}

\section{Introduction}
While deep learning has demonstrated remarkable performance on medical imaging tasks over the past few years, the performance drop usually observed when generalizing from internal to external test data remains a key challenge in the medical application of machine learning models. Supervised learning assumes that training and testing data are sampled from the same distribution, \ie, in-distribution, whereas in practice, the training and testing data typically originate from related domains, but which follow different distributions, \ie, out-of-distribution. This phenomenon, known as domain shift \cite{Quinonero-Candela2009-ml}, hampers the clinical applicability of such models, especially when the annotated datasets are limited in size or the target domain is highly heterogeneous.

One approach to tackling this domain shift problem is domain adaptation, which learns to align the feature distribution of the source domain with that of the target domain in a domain-invariant feature space. However, domain adaptation typically requires access to at least a few data samples from the target domain during training, which is not always available for medical applications. Another approach is domain generalization, which aims to adapt from multiple labeled source domains to an unseen target domain without needing to access data samples from the target domain. However, domain generalization typically requires multi-source training setting. Additionally, these approaches assume the target data are homogeneously sampled from the same distribution, an unrealistic scenario in most real-world medical applications, where models must deal with mixed-domain data (\eg, scanner, protocols, medical sites) without their domain labels. In the present study, our focus is to address a challenging yet practical problem of knowledge transfer from one labeled source domain to multiple target domains, a task referred to as domain agnostic learning \cite{Peng2019-cj} or single-domain generalization \cite{Qiao_2020_CVPR}, where we train the model on source data from a single domain and generalize it to unseen target data from multiple domains. A solution to domain-agnostic learning/single-domain generalization should learn domain-invariant and class-specific visual representations, as humans do.

\begin{figure*}[t]
    \centering
    \includegraphics[width=\textwidth]{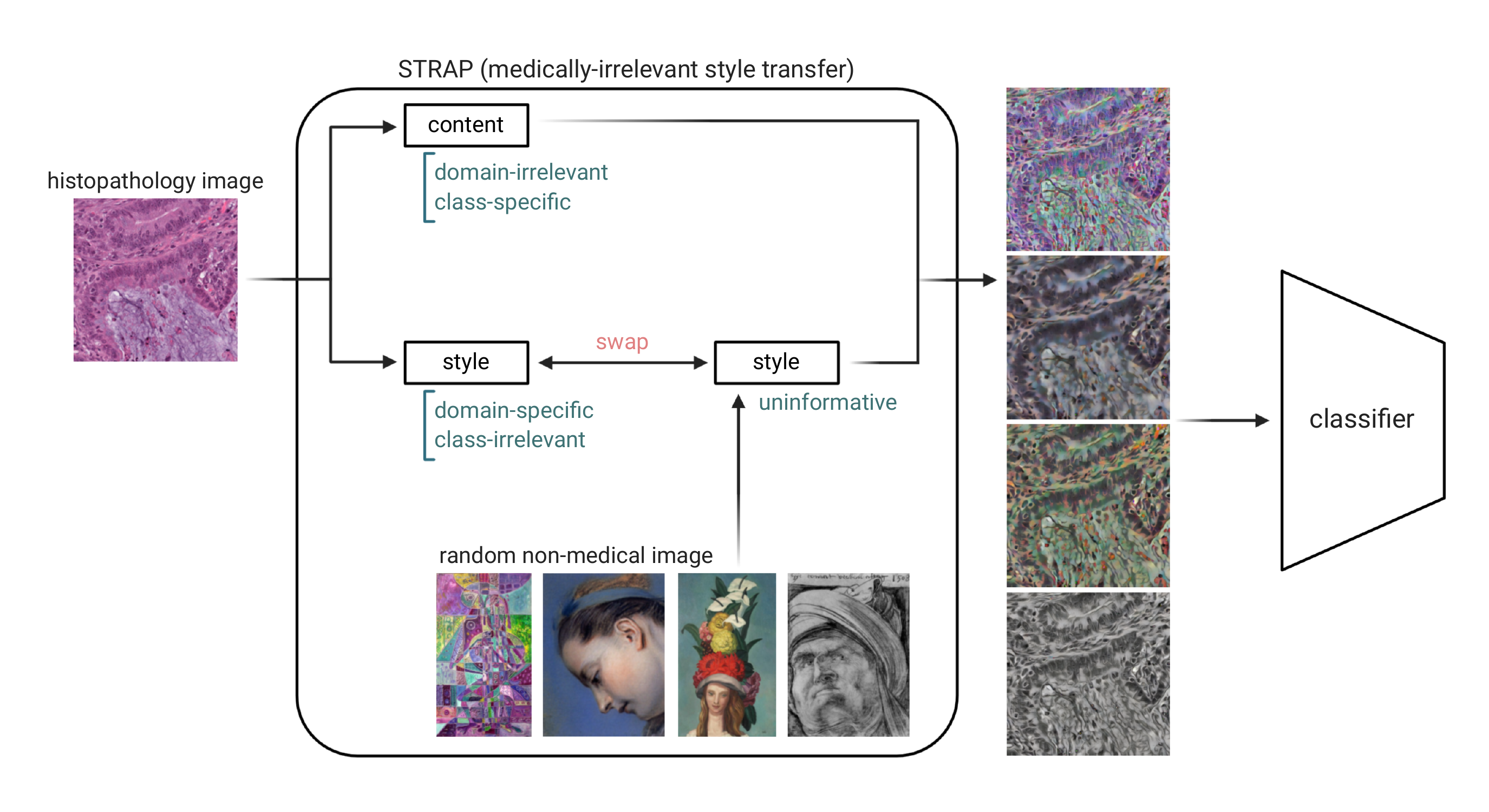}
    \caption{Overview of STRAP}
    \label{fig:overview}
\end{figure*}

Geirhos \etal~\cite{Geirhos2018-xg} showed that 1) convolutional neural networks (CNNs) trained on the ImageNet dataset are biased towards texture, whereas humans are more reliant on global shape for distinguishing classes, 2) CNNs tend not to cope well with domain shifts, \ie, the change in image statistics from those on which the networks have been trained to those which the networks have never seen before, and 3) increasing shape bias by training on a stylized version of the ImageNet generated using style transfer improves accuracy, robustness, and generalizability. 

Neural style transfer~\cite{Gatys2016-ve} refers to a CNN-based image transformation algorithm that manipulates the low-level texture representation of an image, \ie, style, while preserving its semantic content. The original method by Gatys \etal\space uses Gram matrices of the activations from different layers of a CNN to represent the style of an image. Then it uses an iterative optimization method to generate a new image from white noise by matching the activations with the content image and the Gram matrices with the style image. Huang and Belongie later proposed an improved approach called adaptive instance normalization (AdaIN) \cite{Huang2017-tj}, which aligns the mean and variance of the content features with those of the style features. AdaIN enables arbitrary style transfer in real-time. Jackson \etal~\cite{Jackson2019-ie} demonstrated that, in computer vision tasks for natural images, data augmentation via style transfer with randomly selected artistic paintings as a style source improves robustness to domain shift, and can be used as a simple, domain-agnostic alternative to domain adaptation.

In medical imaging, machine learning models often suffer from domain shift in test data caused by heterogeneity from various sources, such as scanners, protocols, and medical sites. We know that human experts, such as radiologists and pathologists, are able to learn domain-agnostic visual representations and, thus, generalize across domains, particularly in the presence of domain shifts. We postulate that 1) human experts in medical imaging are also biased towards shape rather than texture as Geirhos \etal\space demonstrated~\cite{Geirhos2018-xg}, and 2) the low-level texture content of an image tends to be domain-specific, leading to suboptimal performance of deep learning models on domain-shifted unseen data, whereas high-level semantic content is more domain-invariant, from which ubiquitous class-specific visual representations can be learned.

\begin{figure*}[t]
    \centering
    \includegraphics[width=\textwidth]{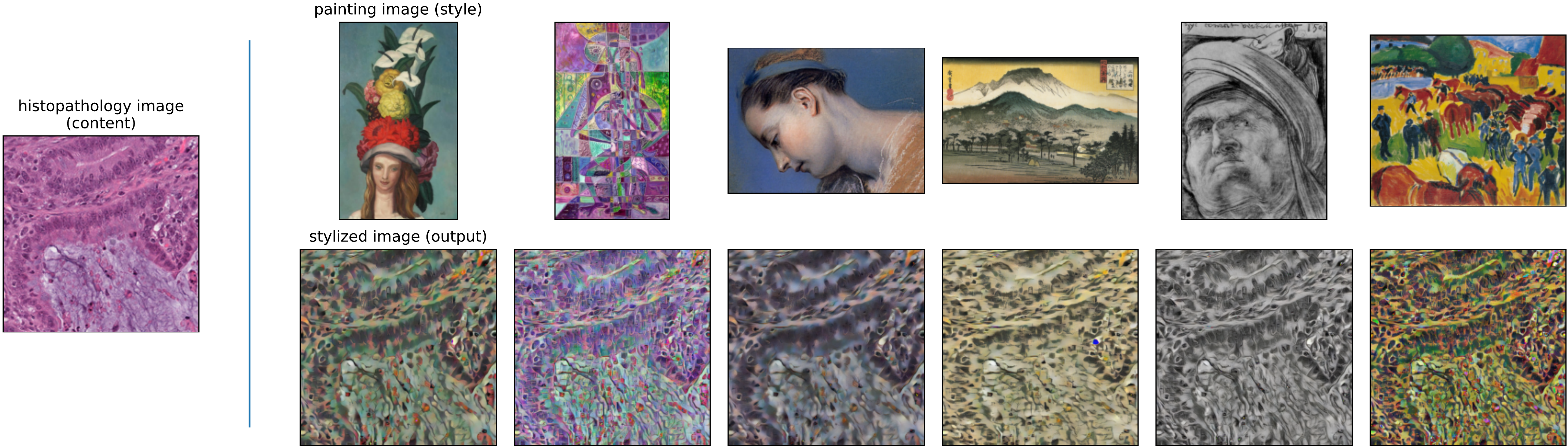}
    \caption{Style transfer with artistic paintings as a style source (stylization coefficient of $1.0$) applied to a histopathology image (content on the left). Overall geometry is preserved, but the style, including texture, color, and contrast, is replaced with an uninformative style of a randomly selected artistic painting.}
    \label{fig:strap}
\end{figure*}

Here, we propose \textbf{STRAP} (\textbf{S}tyle \textbf{TR}ansfer \textbf{A}ugmentation for histo\textbf{P}athology), a form of data augmentation based on random style transfer with non-medical style source, as a solution to learning domain-agnostic visual representation, particularly in computational pathology (Figs. \ref{fig:overview}, \ref{fig:strap}). In this study, the term “domain” refers to scanners, stain and scan protocols, and, more broadly, medical sites. We introduce STRAP as a solution to domain agnostic learning (\ie, single-domain generalization), and then, further assess its efficacy on conventional domain generalization setting (\ie, multi-source domain generalization). More specifically, we studied the proposed approach on two classification tasks in different domain generalization scenarios. The first is classifying colorectal cancer into two distinct genetic subtypes based on microsatellite status using hematoxylin and eosin (H\&E)-stained, formalin-fixed, paraffin embedded (FFPE) whole-slide images (WSIs) of surgically resected colorectal cancers in single-domain generalization setting (models are trained on a single-domain dataset and tested on a mixed-domain dataset), hereafter referred to as genetic subtype classification task. The second is classifying presence or absence of breast cancer metastases in image patches extracted from histopathlogic scans of lymph node sections in multi-source domain generalization setting (models are trained on a multi-source domain dataset and tested on a single-domain dataset), hereafter referred to as tumor identification task. We compare STRAP against two standard baseline methods widely used in computational pathology, stain normalization~\cite{Macenko2009-lb} and stain augmentation~\cite{Tellez2019-uf}, both of which apply medically-relevant transformation to the source images, whereas STRAP performs medically-irrelevant transformation. 

We studied the effect of difference in style source by using artistic paintings, natural imaging, and histopathologic imaging as style sources (the former two apply medically-irrelevant style transfer, whereas the latter applies medically-relevant style transfer), and the effect of difference in stylization coefficient on the STRAP performance. Moreover, to gain insights into the differences in learning dynamics among the three approaches (STRAP, stain normalization, and stain augmentation), we performed following three experiments on the genetic subtype classification task: 1) we tested model performance on stylized version of the out-of-distribution test data; 2) we evaluated differential responses to the low-frequency components of the out-of-distribution test data; and 3) we visualized saliency maps on the low-frequency components of the out-of-distribution test data using integrated gradients~\cite{Sundararajan2017-kt}. The latter two experiments were inspired by Wang \etal~\cite{Wang2020-jq}, who showed that 1) CNNs can exploit high-frequency image components which humans do not consciously perceive and 2) models which exploit low-frequency components generalize better than those which exploit the high-frequency spectrum. 

Our contributions are summarized as follows: 1) we present STRAP, a form of medically-irrelevant data augmentation based on random style transfer for computational pathology; 2) we utilize STRAP to improve both single-domain and multi-source domain generalization for two classification tasks in computational pathology; and 3) our experiments suggest that STRAP helps models learn from semantic contents and low-frequency components of the data, on which humans tend to rely in recognizing objects~\cite{Awasthi2011-mf}.

\section{Methods}

\subsection{Style transfer augmentation with non-medical style source (STRAP)}
\label{subsec:strap}

Inspired by Geirhos \etal~\cite{Geirhos2018-xg} and Jackson \etal~\cite{Jackson2019-ie}, we propose STRAP, a form of medically-irrelevant data augmentation based on random style transfer for computational pathology, which replaces the style of the histopathology image (including texture, color, and contrast) with an uninformative style of a randomly selected non-medical image, while predominantly preserving the semantic content (global object shapes) of the image. We hypothesize that the style of the histopathology images is domain-specific and class-irrelevant, whereas the semantic content is domain-irrelevant and class-specific; therefore, STRAP facilitates learning domain-agnostic representations.

\begin{figure*}[ht!]
    \centering
    \includegraphics[width=\textwidth]{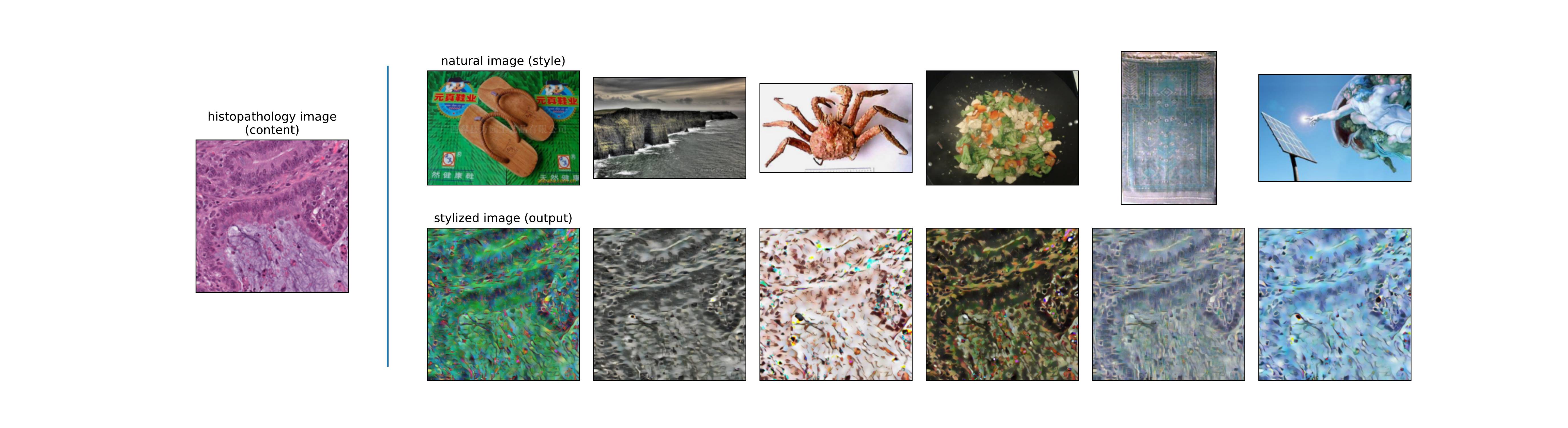}
    \caption{Style transfer with the Natural Imaging style source applied to a histopathology image (content on the left). Overall geometry is preserved, but the style, including texture, color, and contrast, is replaced with the uninformative style of a randomly selected natural image. The outputs are medically irrelevant and resemble the outputs using the Artistic Paintings style source.}
    \label{fig:natural}
\end{figure*}

\begin{figure*}[ht!]
    \centering
    \includegraphics[width=\textwidth]{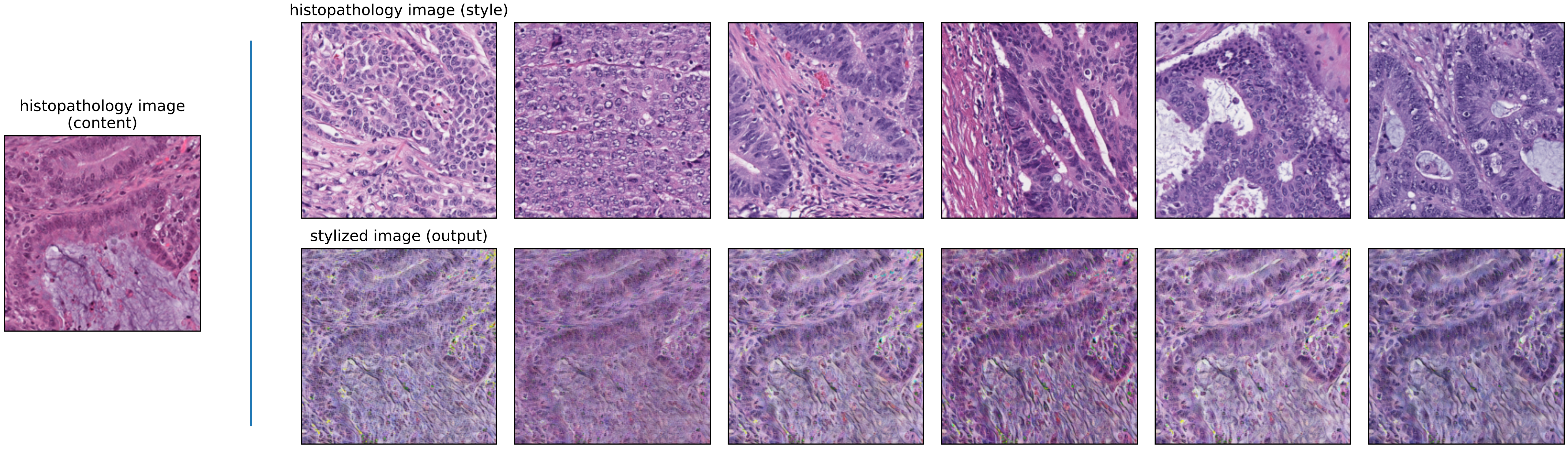}
    \caption{Style transfer with randomly selected histopathologic images from the non-stain normalized version of the Stanford-CRC dataset, applied to a histopathology image (content on the left). The outputs are medically relevant and resemble the outputs obtained with stain augmentation (Figure \ref{fig:stainaug}).}
    \label{fig:histo}
\end{figure*}


We constructed stylized version of the datasets by applying AdaIn style transfer~\cite{Huang2017-tj} following the method proposed in~\cite{Geirhos2018-xg}. AdaIn style transfer takes a content image and an arbitrary style image as inputs, and synthesizes an output image that recombines the content of the former and the style of the latter. After encoding the content and style images in feature space via an encoder, both feature maps are fed to an AdaIN layer that aligns the mean and variance of the content feature maps to those of the style feature maps, producing the target feature maps. Then the stylized output image is generated by a decoder from the target feature maps. We chose AdaIN style transfer because it enables to transfer arbitrary styles in real-time. Each histopathology image was stylized with the style of a randomly selected image from the style source through AdaIN with a stylization coefficient of $1.0$. We studied three distinct style sources: 1) artistic paintings from the Kaggle’s \texttt{Painter by Numbers} dataset ($79,433$ paintings), downloaded via \url{https://www.kaggle.com/c/painter-by-numbers}, hereafter referred to as the Artistic Paintings style source; 2) natural images from the \texttt{miniImageNet} dataset proposed by Vinyals \etal~\cite{Vinyals2016-yv}, consisting of $60,000$ color images from ImageNet with $100$ classes, each having $600$ examples, hereafter referred to as the Natural Imaging style source; and 3) the original Stanford-CRC dataset, containing $66,578$ histopathological images without stain normalization (to preserve the original variability in staining) as described in section \ref{subsubsec:taskone}, hereafter referred to as Histopathologic Imaging style source. The former two apply medically-irrelevant transformation (Figs. \ref{fig:strap} and \ref{fig:natural}), whereas the latter applies medically-relevant transformation (Fig. \ref{fig:histo}). Of note, when applying STRAP, we resized the content histopathology images to $1024\times1024$ pixels and the style source images to $256\times256$ pixels to maintain geometric features of the content images during the stylization. We prepared stylized version of the datasets in advance, because random style transfer via AdaIN as an on-the-fly data augmentation is still computationally expensive.

\begin{figure}[h]
    \centering
    \includegraphics[width=0.7\linewidth]{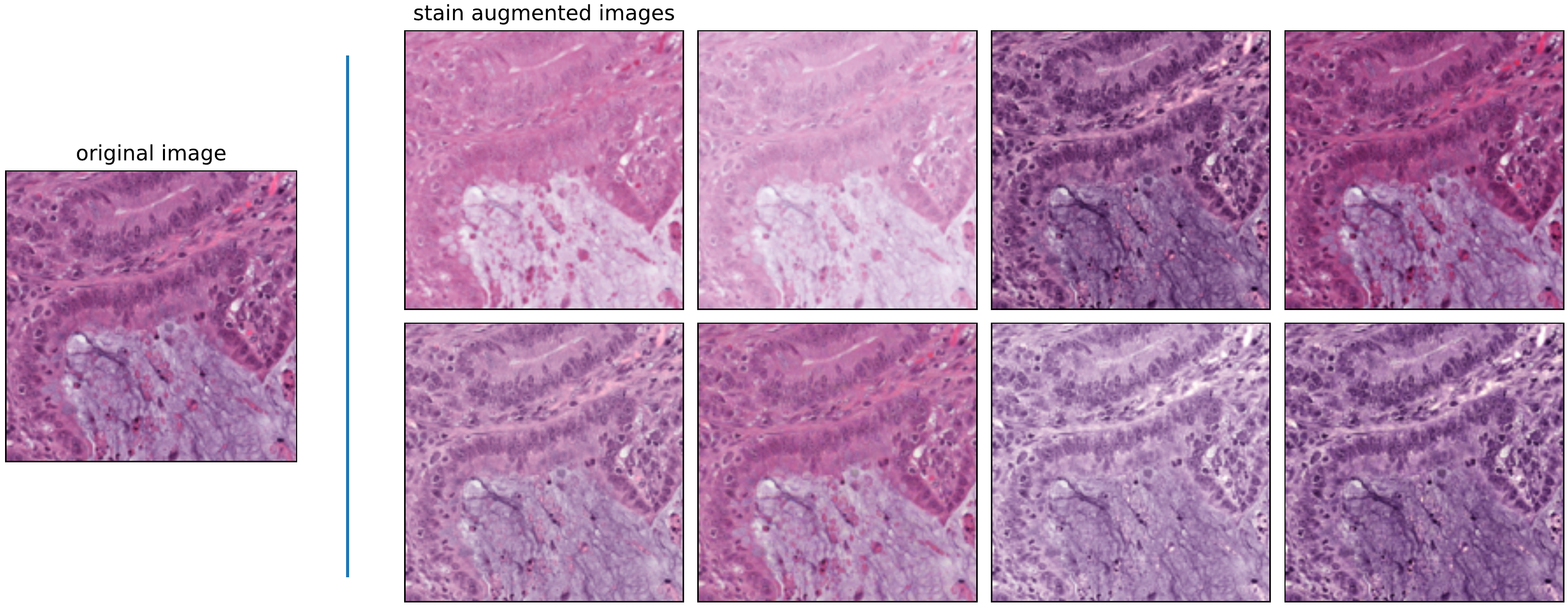}
    \caption{Stain augmentation applied to a histopathology image (original on the left).}
    \label{fig:stainaug}
\end{figure}

We compared STRAP against two standard baseline approaches; stain normalization (SN) and stain augmentation (SA). The STRAP model was trained on stylized datasets alone, whereas the SN model was trained on non-stylized original datasets that were stain-normalized by the Macenko's method~\cite{Macenko2009-lb} and the SA model was trained on non-stylized original datasets with on-the-fly stain augmentation by following the method described by Tellez \etal~\cite{Tellez2019-uf} (Fig. \ref{fig:stainaug}). Stain normalization is a widely used method in computational pathology to account for variations in H\&E staining \cite{Echle2020-pd, Kather2019-fx, Yamashita2021-yz, Komura2018-ia}. On the other hand, Tellez \etal~\cite{Tellez2019-uf} demonstrated that stain augmentation improved classification performance when compared to stain normalization, by increasing the CNN’s ability to generalize to unseen stain variations.

\subsection{Experiments}
\label{subsec:experiments}

We evaluated our proposed approach on two classification tasks, genetic subtype classification and tumor identification, in different domain generalization scenarios, single-domain generalization and multi-source domain generalization, respectively. 

\medskip

\subsubsection{Genetic subtype classification in single-domain generalization setting}
\label{subsubsec:taskone}

The genetic subtype classification task was to classify colorectal cancer into two distinct genetic subtypes based on microsatellite status (either microsatellite stable (MSS) or microsatellite unstable (MSI)) using hematoxylin and eosin (H\&E)-stained, formalin-fixed, paraffin embedded (FFPE) whole-slide images (WSIs) of surgically resected colorectal cancers. We evaluated our proposed approach in single-domain generalization setting (models are trained on a single-domain dataset and tested on a mixed-domain dataset).

We reused three datasets that were created and used in previous publications; Stanford-CRC from Yamashita \etal~\cite{Yamashita2021-yz} and CRC-DX-TRAIN as well as CRC-DX-TEST from Kather \etal~\cite{Kather2019-fh} (See the original publications for details, such as inclusion/exclusion criteria and clinico-pathological parameters). These datasets consists of image patches called tiles, which were generated from the WSIs with a size of $512\times512$ pixels at a resolution of $0.5$ \si{\micro\metre}/pixel and subsequently stain normalized with the Macenko's method~\cite{Macenko2009-lb}. 

The Stanford-CRC is an in-house dataset that originates from a single institution and contains $66,578$ image tiles ($31,789$ tiles from $50$ MSS and $34,789$ tiles from $50$ MSI H\&E-stained FFPE WSI) from $100$ unique patients. The WSIs were originally scanned at $40\times$ base magnification level ($0.25$ \si{\micro\metre}/pixel). This single-institutional dataset has equal class distribution, with $50$ MSS and $50$ MSI patients.

The CRC-DX-TRAIN dataset stems from the TCGA-COAD and TCGA-READ diagnostic slide collections of the Cancer Genome Atlas (TCGA)~\cite{Muzny2012-br}, consisting of data from $18$ institutions with various scanners and protocols, \ie, a multi-domain dataset, and contains $93,408$ image tiles ($46,704$ tiles from $223$ MSS and $46,704$ tiles from $40$ MSI H\&E-stained FFPE WSI) from $263$ unique patients. The WSI were scanned at either $20\times$ or $40\times$ base magnification ($0.5$ or $0.25$ \si{\micro\metre}/pixel). This multi-institutional dataset was balanced in class distribution.

The CRC-DX-TEST dataset stems from the same diagnostic slide collections of TCGA as the CRC-DX-TRAIN dataset, \ie, consisting of data from $18$ institutions with various scanners and protocols, and contains $99,904$ image tiles ($70,569$ tiles from $74$ MSS and $29,335$ tiles from $26$ MSI H\&E-stained FFPE WSI) from $100$ unique patients. This multi-institutional dataset maintains the original class imbalance, which reflects real-world prevalence of MSI in colorectal cancer.

We performed both in-distribution and out-of-distribution experiments using the above three datasets. For in-distribution analysis, models were trained on CRC-DX-TRAIN and evaluated on CRC-DX-TEST. Our out-of-distribution experiment follows the single-domain generalization setting, where models were trained on single-domain Stanford-CRC dataset and evaluated on multi-source domain CRC-DX-TEST dataset. We applied 4-fold cross-validation to account for the selection bias introduced by randomness in splitting Stanford-CRC, given its relatively limited sample size; therefore, average and standard deviation of the evaluation metric across the folds were reported. Of note, all the STRAP models were trained on the stylized version of the training datasets by applying the style transfer method described in section \ref{subsec:strap}.

We employed the MobileNetv2~\cite{Sandler2018-hl} model pretrained on ImageNet~\cite{Russakovsky2015-py} via transfer learning with stochastic gradient descent with momentum~\cite{Qian1999-zu}, using a fixed learning rate of $4\mathrm{e}{-3}$ and epoch of $40$, along with early stopping with a patience of five. We used a binary cross entropy loss. All input images were resized to $224\times224$ pixels before being fed into the network. Random horizontal and vertical flipping (with a probability of $0.5$ for each) and random resized cropping were applied as a common data augmentation method. Tile-wise model outputs were aggregated into a patient-wise score by taking their average. The particular metric of interest was the area under the receiver-operating-characteristic curve (AUROC).

We further compared the STRAP model against two state-of-the-arts, Kather \etal~\cite{Kather2019-fx} and Yamashita \etal~\cite{Yamashita2021-yz} in the same single-domain generalization scenario for genetic subtype classification. Both approaches are similar to our SN baseline, though there are some differences in model architecture, training protocols, and configuration of data augmentation. For example, Kather \etal~used a ResNet-18 architecture~\cite{7780459} and applied horizontal and vertical flips and random translation along the $x$ and $y$ axes for data augmentation. Similarly, Yamashita \etal~used a MobileNetV2 architecture and applied data augmentation with random horizontal flips, random rotations, and random color jitter. Model performance for Kather \etal~\cite{Kather2019-fx} and Yamashita \etal~\cite{Yamashita2021-yz} was either computed using the code available at \url{https://github.com/jnkather/MSIfromHE} and \url{https://github.com/rikiyay/MSINet}, respectively, or obtained from the literature.

\smallskip

\textbf{Impact of differences in style source and stylization coefficient}
As sensitivity analyses, we performed two additional experiments. First, we studied the effect of difference between medically-irrelevant and medically-relevant STRAP approaches. More specifically, we compared the performance of the STRAP models using Artistic Paintings and Natural Imaging style sources (medically-irrelevant approach) against the STRAP model with Histopathologic Imaging style source (medically-relevant approach) on the genetic subtype classification task. We also studied the effect of difference in stylization coefficient, where the STRAP models using stylization coefficient of $1.0$, $0.8$, and $0.6$ were compared on the genetic subtype classification task.


\smallskip

\textbf{Assessment on stylized images with random test-time styles}
To understand how content and style are being utilized, we compared the model performance for STRAP, SA, and SN on stylized version of the CRC-DX-TEST with random test-time styles of the Natural Imaging style source. For STRAP, we tested both STRAP with Artistic Paintings and STRAP with Histopathologic Imaging to assess the difference in sensitivity between medically-irrelevant and medically-relevant approaches.

\smallskip

\begin{figure}[h]
    \centering
    \includegraphics[width=0.65\linewidth]{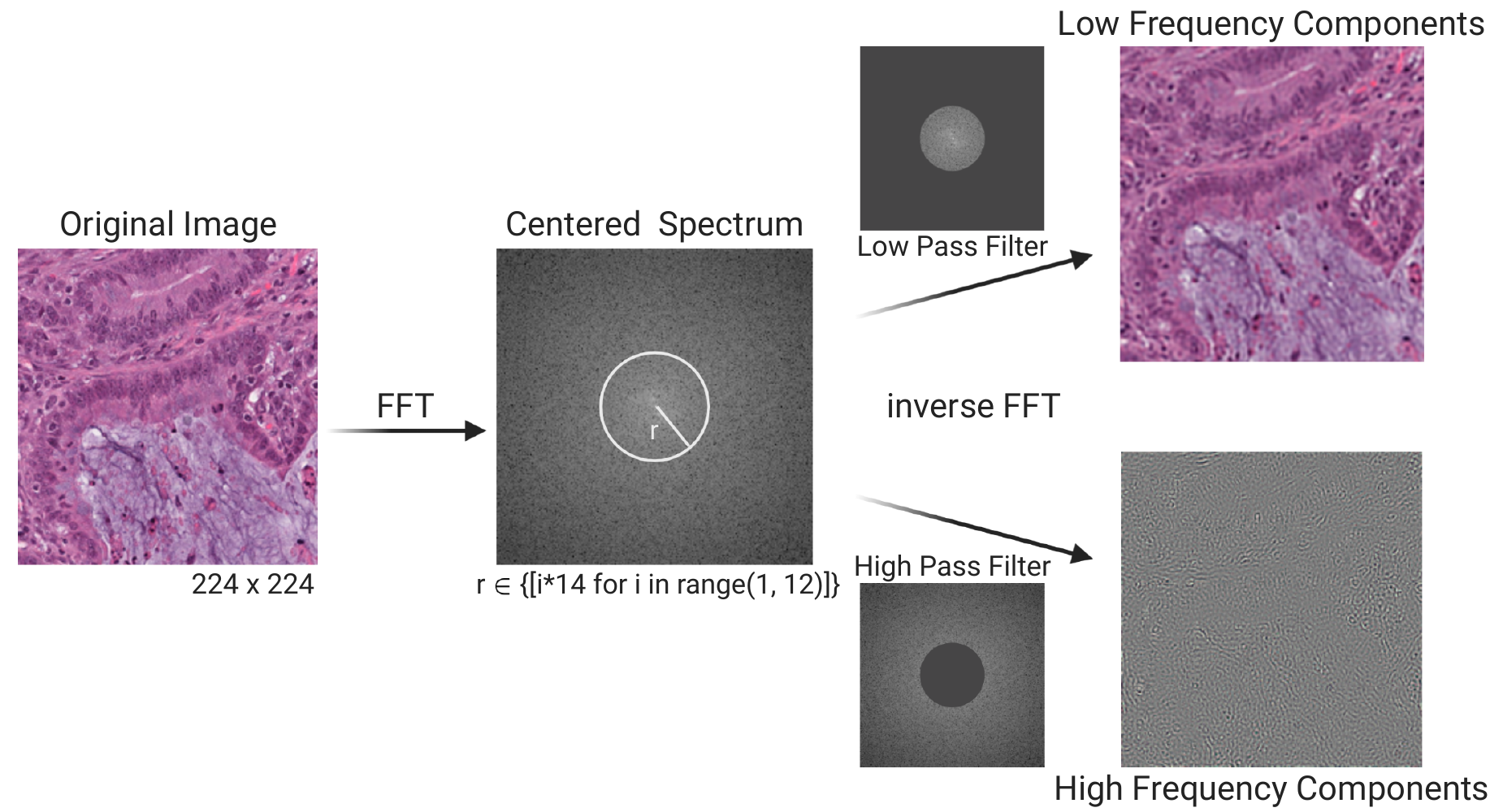}
    \caption{A schema for generating low-frequency components of an image. Image tiles are decomposed into low- and high-frequency components by applying the fast Fourier transform (FFT) algorithm. Low-frequency components are extracted from the centralized frequency spectrum by applying circular low-pass filters with various radii. All frequencies outside the circle were set to zero and the inverse FFT was applied subsequently. Of note, the high frequency components were not used in this study.}
    \label{fig:lowfreq}
\end{figure}

\textbf{Assessment on low-frequency components}
To gain insights into what frequency components the three models (STRAP, SA, and SN) exploit for learning representations, we tested model performance on the low-frequency components of the CRC-DX-TEST dataset, hereafter referred to as LF-CRC-DX-TEST. We constructed the LF-CRC-DX-TEST dataset by following the method described in~\cite{Wang2020-jq}, where all image tiles in the CRC-DX-TEST dataset were decomposed into low- and high-frequency components by applying the fast Fourier transform (FFT) algorithm. Low-frequency components were obtained from the centralized frequency spectrum by applying circular low-pass filters with various radii. All frequencies outside the circular filter were set to zero and the inverse FFT was applied subsequently to get the low-frequency images (Fig. \ref{fig:lowfreq}). To identify the low-pass filter size that corresponds to the highest model performance, the AUROC for each of the STRAP, SA, and SN models was assessed using varying low-pass filter sizes (the radii ranged from 14 to 154).

We also visualized saliency maps on the LF-CRC-DX-TEST (with a low-pass filter size of 70) using integrated gradients attributions~\cite{Sundararajan2017-kt} to highlight which pixels of an input image contribute more to model inference.

\medskip

\subsubsection{Tumor identification in multi-domain generalization setting}
\label{subsubsec:tasktwo}

The tumor identification task was to classify presence or absence of breast cancer metastases in image patches extracted from histopathlogic scans of lymph node sections in multi-source domain generalization setting (models are trained on a multi-source domain dataset and tested on a single-domain dataset). 

We used the CAMELYON17-WILDS dataset~\cite{wilds2021}, a patch-based variant of the original Camelyon17 dataset~\cite{bandi2018detection} created as a benchmark dataset for domain generalization, where the domains are hospitals and the goal is to learn models that generalize to data from a hospital that is not in the training subset. The specific task is to predict if a given region of tissue contains any tumor tissue, which was modeled as binary classification, where the input is a $96\times96$-pexel histopathological image, the label is a binary indicator of whether the central $32\times32$ region contains any tumor tissue.

The CAMELYON17-WILDS dataset was adapted from WSIs of breast cancer metastases in lymph nodes sections, obtained from the CAMELYON17 challenge~\cite{bandi2018detection}, where the WSIs were scanned at a resolution of $0.23$–$0.25$\si{\micro\metre}, and each WSI contains multiple resolution levels, with approximately $10,000\times20,000$ pixels at the highest resolution level. Image patches were generated using the third-highest resolution level, corresponding to reducing the size of each dimension by a factor of 4. The CAMELYON17-WILDS dataset comprises $455,954$ patches extracted from $50$ WSIs of breast cancer metastases in lymph node sections, with $10$ WSIs from each of 5 hospitals. The label for each patch was determined by the segmentation masks manually annotated with tumor regions by pathologists, which were provided along with the original Camelyon17 dataset. We split the CAMELYON17-WILDS dataset by domain (\ie, which hospital the patches were taken from) using the metadata. We used the \texttt{Test(OOD)} subset of the CAMELYON17-WILDS dataset as our out-of-distribution test subset, which contains $85,054$ patches taken from $10$ WSIs from the 5th hospital (\texttt{center} 2 in the provided metadata), which was chosen by the original WILDS project because its patches were the most visually distinctive. We split the rest patches into training and validation based on the \texttt{split} column provided in the metadata (\texttt{split} 0 for training and \texttt{split} 1 for validation), where $333,866$ and $37,034$ patches taken from $40$ WSIs, with $10$ WSIs from each of the 4 hospitals, were assigned to the training and validation sets, respectively. Of note, the training/validation and test sets comprise class-balanced patches from separate hospitals (See the original publication~\cite{wilds2021} for more details).

We employed the ResNet-50~\cite{7780459} model pretrained on ImageNet~\cite{Russakovsky2015-py} via transfer learning with stochastic gradient descent with momentum~\cite{Qian1999-zu}, using a fixed learning rate of $4\mathrm{e}{-3}$ and epoch of $40$, along with early stopping with a patience of five. We used a binary cross entropy loss. Random horizontal and vertical flipping (with a probability of $0.5$ for each) and random resized cropping were applied as a common data augmentation method. Model performance was evaluated by average accuracy and AUROC across patches. Of note, unlike the genetic subtype classification task where the ground truth labels are patient-level, the ground truth labels for the tumor identification task are patch-level, meaning no output aggregation procedure is required.

\subsection{Statistical analysis}
\label{subsec:stats}

We assessed model performance using the AUROC for genetic subtype classification, and accuracy as well as AUROC for tumor identification. We calculated $95\%$ confidence intervals (CI) using bootstrapping with the percentile method with $2,000$ resamples. Statistical comparisons were performed using the DeLong's test~\cite{10.2307/2531595} for individual AUROC, a paired t-test for average AUROC, and a permutation test with $2,000$ resamples for accuracy. For the main analyses of both genetic subtype classification and tumor identification (results are shown in Tables \ref{tab:r_msi} and \ref{tab:r_wilds}, respectively), p-values were adjusted using the Benjamini-Hochberg method~\cite{Benjamini1995-hx} to account for multiple comparisons by controlling the false positive rate to less than $0.10$. Otherwise, a two-tailed alpha criterion of $0.05$ was used for statistical significance.

\section{Experimental Results}

\subsection{Genetic subtype classification in single-domain generalization setting}
\label{subsec:r_taskone}

The STRAP model with Artistic Paintings style source achieved an average AUROC of $0.876$ on the out-of-distribution multi-domain CRC-DX-TEST dataset, and outperformed the SA, SN, and the two state-of-the-art models (Table \ref{tab:r_msi}). STRAP also demonstrated a minimal, even negative, performance drop from in-distribution to out-of-distribution testing (see column \texttt{Delta} in Table \ref{tab:r_msi}), whereas SA presented near-zero performance drop and the others showed positive performance drops. These results suggest that the STRAP model has the ability to learn more discriminative and generalizable (\ie, class-specific and domain-irrelevant) visual representations, compared to the other approaches that may exploit some extent of the domain-specific features.

\begin{table}[h]
  \centering
  \caption{Comparison of style transfer augmentation (STRAP), stain augmentation (SA), stain normalization (SN), and two state-of-the-arts on in-distribution and out-of-distribution (single-domain generalization) scenarios on the genetic subtype classification task.}
  \begin{threeparttable}
  \small
  \begin{tabular*}{\textwidth}{cccccc}
    \toprule
    & \multicolumn{2}{c}{CRC-DX-TRAIN $\rightarrow$ CRC-DX-TEST (ID)} & \multicolumn{2}{c}{Stanford-CRC $\rightarrow$ CRC-DX-TEST (OOD)} & \multirow{2}{*}{\begin{tabular}[c]{@{}c@{}}Delta\S\\ (ID$-$OOD)\end{tabular}} \\
    & AUROC\dag & p-value (vs STRAP) & AUROC\ddag & p-value (vs STRAP) & \\
    \midrule
    STRAP (AP) & \bf{0.847 {[}0.741, 0.932{]}} & REF & \bf{0.876 (0.015)} & REF & \bf{$-$0.029} \\
    SA & 0.816 {[}0.709, 0.917{]} & 0.471 & 0.814 (0.020) & 0.002\textasteriskcentered & 0.002 \\
    SN & 0.794 {[}0.684, 0.892{]} & 0.456 & 0.765 (0.031) & 0.003\textasteriskcentered & 0.029 \\
    Kather \etal & 0.759 {[}0.632, 0.873{]} & 0.219 & 0.742 (0.013) & 0.001\textasteriskcentered & 0.018 \\
    Yamashita \etal & 0.816 {[}0.712, 0.914{]} & 0.456 & 0.786 (0.020) & 0.010\textasteriskcentered & 0.030 \\
    \bottomrule
  \end{tabular*}
  \begin{tablenotes}
  \item Arrows indicate: train data $\rightarrow$ test data, \eg, CRC-DX-TRAIN $\rightarrow$ CRC-DX-TEST means training on CRC-DX-TRAIN and testing on CRC-DX-TEST.
  \item \textasteriskcentered\space indicates a significant difference.
  \item \dag\space represents AUROC with $95\%$ CI in square brackets.
  \item \ddag\space represents average AUROC of models obtained via cross-validation, with standard deviation in parentheses.
  \item \S\space indicates average performance drop from in-distribution (CRC-DX-TRAIN $\rightarrow$ CRC-DX-TEST) to out-of-distribution (Stanford-CRC $\rightarrow$ CRC-DX-TEST) scenarios.
  \item Stylization coefficient (alpha) of $1.0$ was used for the STRAP model.
  \item P-values were adjusted using the Benjamini-Hochberg method~\cite{Benjamini1995-hx}.
  \item Abbreviations: AP, Artistic Paintings; AUROC, areas under the receiver-operating-characteristic curve; CV, cross-validation; ID, in-distribution; OOD, out-of-distribution; SA, stain augmentation; SN, style normalization; STRAP, style transfer augmentation.
  \end{tablenotes}
  \end{threeparttable}
  \label{tab:r_msi}
\end{table}

\medskip

\subsubsection{Impact of differences in style source}
\label{subsubsec:r_source}

For genetic subtype classification, medically-irrelevant STRAP using Artistic Paintings and Natural Imaging as style sources achieved superior performance compared to the medically-relevant STRAP using Histopathologic Imaging as style source. In comparison to Histopathologic Imaging, the Artistic Paintingsyielded a significantly higher performance, whereas there was no statistically significant difference between the Natural Imaging and Histopathologic Imaging style sources (Table \ref{tab:r_source}).

\begin{table}[h]
  \centering
  \caption{Effect of different style sources on STRAP model performance.}
  \begin{threeparttable}
  \begin{tabular}{ccc}
    \toprule
    & \multicolumn{2}{c}{Stanford-CRC $\rightarrow$ CRC-DX-TEST} \\
    Style Source & AUROC\dag & p-value (vs HI) \\
    \midrule
    Artistic Paintings (AP) & \bf{0.876 (0.015)} & 0.037\textasteriskcentered \\
    Natural Imaging (NI) & 0.867 (0.016) & 0.088 \\
    Histopathologic Imaging (HI) & 0.822 (0.042) & REF \\
    \bottomrule
  \end{tabular}
  \begin{tablenotes}
  \item Arrow indicates: train data $\rightarrow$ test data, \ie, Stanford-CRC $\rightarrow$ CRC-DX-TEST means training on Stanford-CRC and testing on CRC-DX-TEST.
  \item \textasteriskcentered\space indicates a significant difference.
  \item \dag\space represents average AUROC of models obtained via cross-validation, with standard deviation in parentheses.
  \item Stylization coefficient (alpha) of $1.0$ was used for the STRAP model.
  \item Abbreviations: AUROC, areas under the receiver-operating-characteristic curve; CV, cross-validation.
 \end{tablenotes}
 \end{threeparttable}
 \label{tab:r_source}
\end{table}

\medskip

\subsubsection{Impact of stylization coefficient}
\label{subsubsec:r_alpha}

We also tested the effect of the stylization coefficient on STRAP model performance. We found that, among stylization coefficients of $1.0$, $0.8$, and $0.6$, the larger the stylization coefficient (\ie, with a stylization coefficient of $1.0$), the higher the model performance (Table \ref{tab:r_alpha}), which suggests that the STRAP model can learn more discriminative and generalizable representations when more low-level content within an image was removed and replaced by the style transfer operation.

\begin{table}[h]
  \centering
  \caption{Effect of stylization coefficient on STRAP model performance.}
  \begin{threeparttable}
  \begin{tabular}{ccc}
    \toprule
    & \multicolumn{2}{c}{Stanford-CRC $\rightarrow$ CRC-DX-TEST} \\
    Stylization Coefficient & AUROC\dag & p-value (vs SC 1.0) \\
    \midrule
    SC $1.0$ & \bf{0.876 (0.015)} & REF \\
    SC $0.8$ & 0.856 (0.036) & 0.189 \\
    SC $0.6$ & 0.846 (0.024) & 0.024\textasteriskcentered \\
    \bottomrule
  \end{tabular}
  \begin{tablenotes}
  \item Arrow indicates: train data $\rightarrow$ test data, \ie, Stanford-CRC $\rightarrow$ CRC-DX-TEST means training on Stanford-CRC and testing on CRC-DX-TEST.
  \item \textasteriskcentered\space indicates a significant difference.
  \item \dag\space represents average AUROC of models obtained via cross-validation, with standard deviation in parentheses.
  \item Abbreviations: AUROC, areas under the receiver-operating-characteristic curve; CV, cross-validation; SC, stylization coefficient.
  \end{tablenotes}
  \end{threeparttable}
  \label{tab:r_alpha}
\end{table}

\subsubsection{Assessment on stylized images with random test-time styles}
\label{subsubsec:r_random_style}


We assessed the model performance on stylized version of CRC-DX-TEST created using Natural Imaging as style source. As shown in Table \ref{tab:r_random_style}, STRAP with Artistic Paintings style source, a medically-irrelevant style transfer, achieved significantly higher performance compared to SA and SN and tended to have higher performance compared to medically-relevant STRAP with Histopathologic Imaging style source. STRAP with Artistic Paintings also demonstrated the smallest performance difference between original and stylized CRC-DX-TEST. This result suggests that the medically-irrelevant STRAP successfully biased the networks to content/shape, which may explain its superior performance and out-of-distribution generalizability compared to the other three (\ie, medically-relevant STRAP, SA, and SN) approaches.

\begin{table}[h]
  \centering
  \caption{Model performance on CRC-DX-TEST dataset with and without random test-time styles.}
  \begin{threeparttable}
  \begin{tabular}{ccccc}
    \toprule
    & \multicolumn{2}{c}{AUROC on CRC-DX-TEST\dag} & & \\
    & Original & Stylized & Delta\S & p-value\ddag \\
    \midrule
    STRAP (AP) & \bf{0.876 (0.015)} & \bf{0.830 (0.020)} & \bf{0.046} & REF \\
    STRAP (HI) & 0.822 (0.042) & 0.711 (0.077) & 0.111 & 0.085 \\
    SA & 0.814 (0.020) & 0.726 (0.055) & 0.084 & 0.047\textasteriskcentered \\
    SN & 0.765 (0.031) & 0.633 (0.577) & 0.132 & 0.015\textasteriskcentered \\
    \bottomrule
  \end{tabular}
  \begin{tablenotes}
  \item \textasteriskcentered\space indicates a significant difference.
  \item \dag\space represents average AUROC with standard deviation in parentheses.
  \item \S\space represents average performance difference between original and stylized CRC-DX-TEST.
  \item \ddag\space represents p-value for comparing model perfromance on stylized CRC-DX-TEST.
  \item Stylized CRC-DX-TEST was created using Natural Imaging style source.
  \item Stylization coefficient (alpha) of $1.0$ was used for the STRAP models.
  \item Abbreviations: AP, Artistic Paintings; AUROC, areas under the receiver-operating-characteristic curve; HI, Histopathologic Imaging; SA, stain augmentation; SN, style normalization; STRAP, style transfer augmentation.
 \end{tablenotes}
 \end{threeparttable}
 \label{tab:r_random_style}
\end{table}

\subsubsection{Assessment on low-frequency components}
\label{subsubsec:r_lowfreq}

\begin{figure}[ht!]
    \centering
    \includegraphics[width=0.6\linewidth]{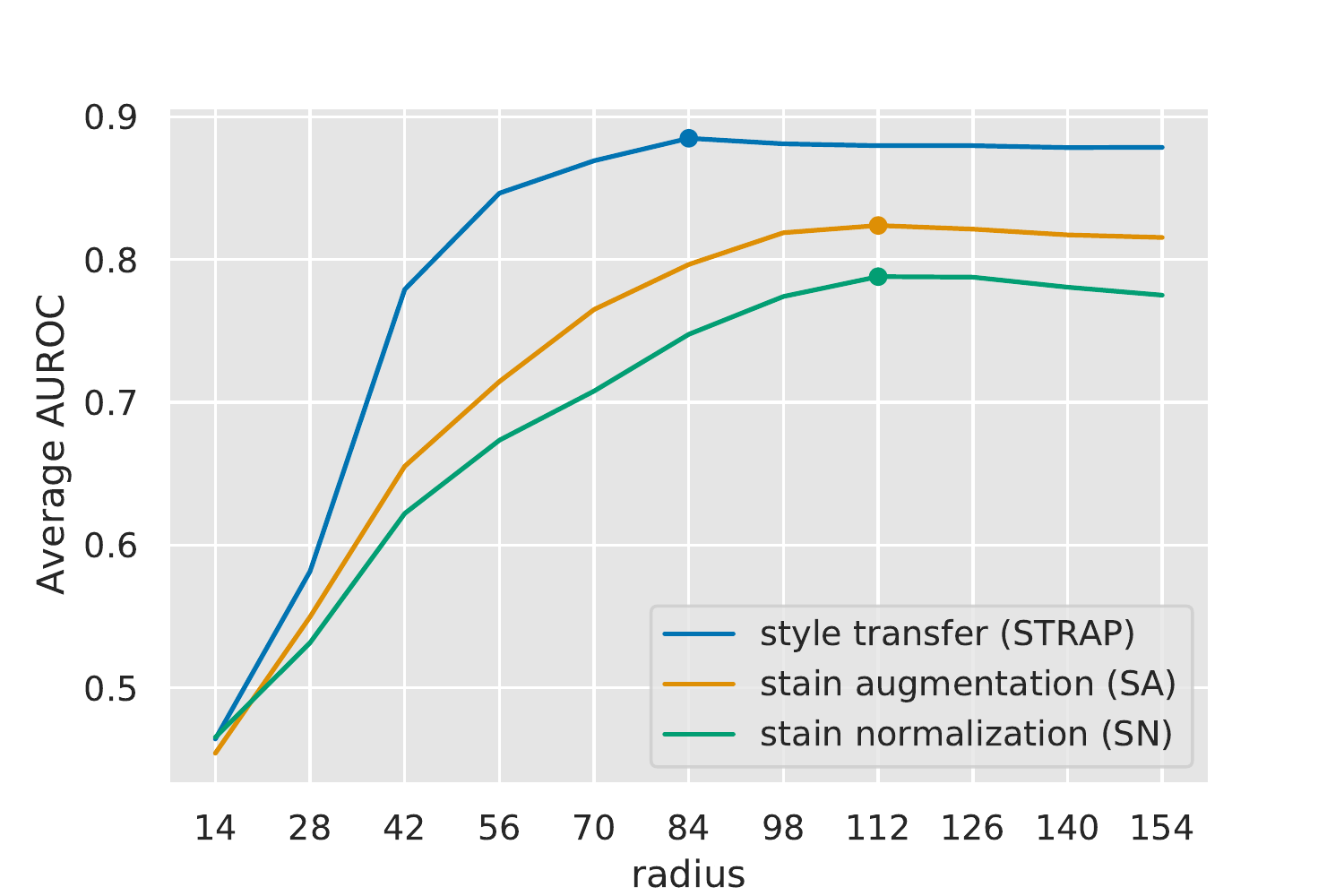}
    \caption{Results of the experiments using the low-frequency components of the CRC-DX-TEST dataset (LF-CRC-DX-TEST). The $x$-axis represents the radii of low-pass filters used to generate the LF-CRC-DX-TEST dataset, and the $y$-axis shows the average area under the receiver-operating-characteristic curves (AUROC) across cross-validation folds. Each dot marker represents the corresponding peak performance.}
    \label{fig:r_lowfreq}
\end{figure}

We evaluated the STRAP, SA, and SN models on the LF-CRC-DX-TEST dataset with a wide range of low-pass filter sizes. As shown in Fig. \ref{fig:r_lowfreq}, the STRAP model reached its peak performance at a radius of $84$, whereas the other two reached their peaks at a radius of $112$. These results suggest that the STRAP model can exploit lower-frequency components for learning representations, whereas the other two baselines rely more on higher-frequency components. We speculate that, because the STRAP model is biased toward shape~\cite{Geirhos2018-xg}, it performs well on lower-frequency components, which preserve most of the geometry and thus, almost look identical to the original image to human. On the contrary, the baseline SA and SN approaches do not address style and content explicitly and thus, require texture and/or higher-frequency components to reach their peak performance.

\begin{figure}[ht!]
    \centering
    \includegraphics[width=0.7\linewidth]{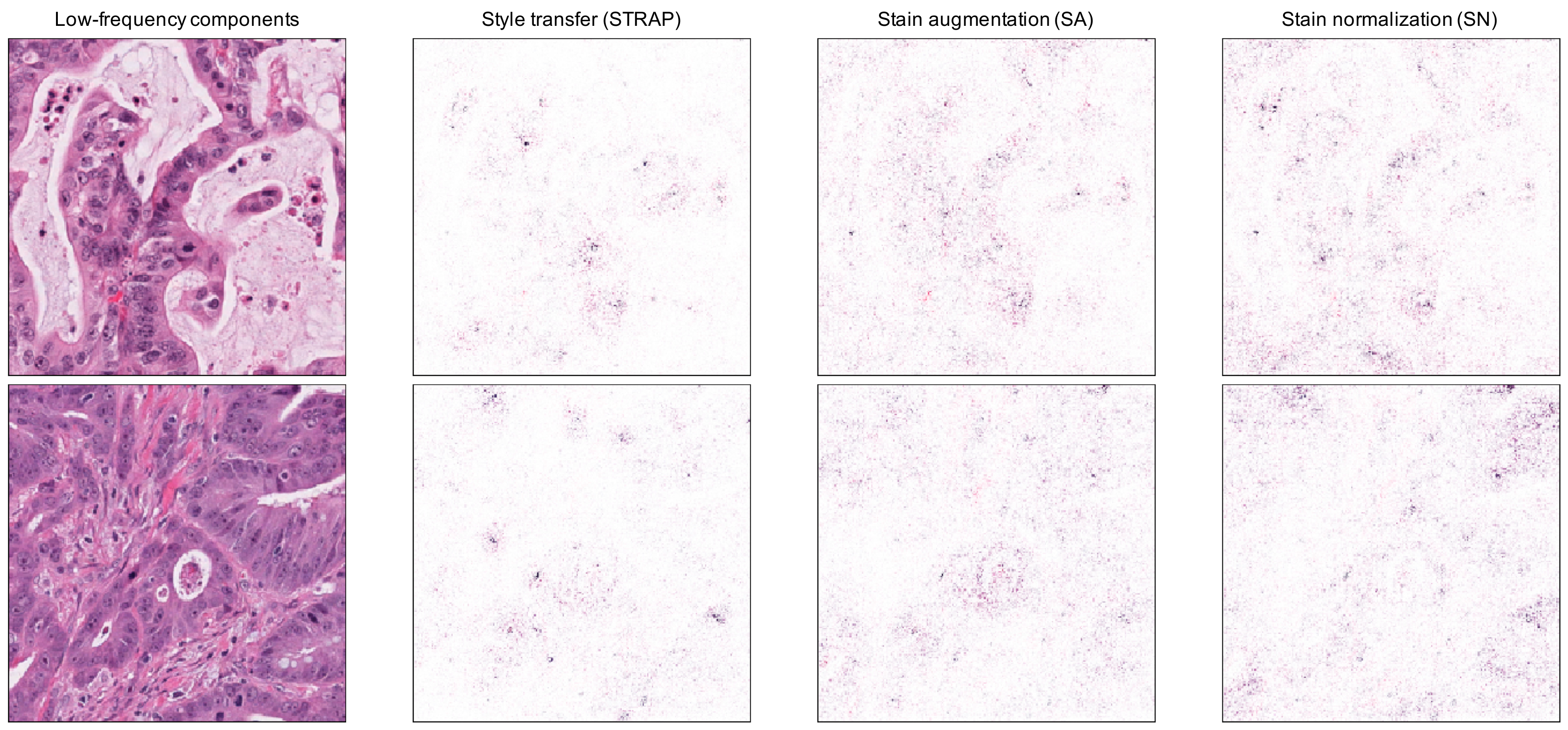}
    \caption{Pixel-wise integrated gradient attributions of the low-frequency components (generated with a radius of $70$) of the CRC-DX-TEST dataset (LF-CRC-DX-TEST), visualized as saliency maps for the STRAP, SA, and SN models.}
    \label{fig:r_saliency}
\end{figure}

Saliency maps with integrated gradients show that the STRAP model presented high attributions at specific areas and less diffusely distributed attributions, whereas the SA and SN models showed more broadly distributed attributions that might correspond to the low-level texture content of the images (Fig. \ref{fig:r_saliency}). A board-certified, subspecialty gastrointestinal pathologist interpreted these saliency maps and concluded that STRAP picks up tumor-infiltrating lymphocytes as well as mitotic figures, which are well-known human-recognizable histomorphologic features that are associated with the genetic subtype of interest.

\subsection{Tumor identification in multi-domain generalization setting}
\label{subsec:r_tasktwo}

On the tumor identification task using CAMELYON17-WILDS dataset, we developed models on the tiles from four out of five hospitals, and assessed the performance on the tiles from the 5th hospital, \ie, the \texttt{Test(OOD)} subset of the CAMELYON17-WILDS (multi-domain generalization setting). As shown in Table \ref{tab:r_wilds}, the medically-irrelevant STRAP model using Artistic Paintings style source achieved the highest accuracy and AUROC with significant differences compared to the other approaches. The results have a similar trend as those for genetic subtype classification, where another medically-irrelevant STRAP with Natural Imaging style source demonstrated the second highest performance, medically-relevant STRAP using Histopathologic Imaging style source and SA presented similar performance that was the next highest, and SN showed the lowest performance.

\begin{table}[h]
  \centering
  \caption{Comparison of style transfer augmentation (STRAP), stain augmentation (SA), stain normalization (SN) on out-of-distribution (multi-source domain generalization) scenarios on the tumor identification task.}
  \begin{threeparttable}
  \begin{tabular}{ccccc}
    \toprule
    & \multicolumn{4}{c}{CAMELYON17-WILDS} \\
    & Accuracy & p-value & AUROC & p-value \\
    \midrule
    STRAP (AP) & \bf{0.937 {[}0.935, 0.938{]}} & REF & \bf{0.981 {[}0.980, 0.982{]}} & REF \\
    STRAP (NI) & 0.923 {[}0.921, 0.925{]} & $<$0.0001\textasteriskcentered & 0.977 {[}0.976, 0.978{]} & $<$0.0001\textasteriskcentered \\
    STRAP (HI) & 0.831 {[}0.829, 0.834{]} & $<$0.0001\textasteriskcentered & 0.888 {[}0.885, 0.890{]} & $<$0.0001\textasteriskcentered \\
    SA & 0.833 {[}0.830, 0.835{]} & $<$0.0001\textasteriskcentered & 0.916 {[}0.914, 0.918{]} & $<$0.0001\textasteriskcentered \\
    SN & 0.631 {[}0.628, 0.634{]} & $<$0.0001\textasteriskcentered & 0.859 {[}0.856, 0.861{]} & $<$0.0001\textasteriskcentered \\
    \bottomrule
  \end{tabular}
  \begin{tablenotes}
  \item \textasteriskcentered\space indicates a significant difference.
  \item Stylization coefficient (alpha) of $1.0$ was used for the STRAP models.
  \item P-values were adjusted using the Benjamini-Hochberg method~\cite{Benjamini1995-hx}.
  \item Abbreviations: AP, Artistic Paintings; AUROC, areas under the receiver-operating-characteristic curve; HI, Histopathologic Imaging; NI, Natural Imaging; SA, stain augmentation; SN, style normalization; STRAP, style transfer augmentation.
 \end{tablenotes}
 \end{threeparttable}
 \label{tab:r_wilds}
\end{table}

\section{Discussion}

We present \textbf{STRAP} (\textbf{S}tyle \textbf{TR}ansfer \textbf{A}ugmentation for histo\textbf{P}athology), which achieved improved performance and generalizability when compared with two standard baselines (stain augmentation (SA) and stain normalization (SN)) on two classification tasks (\ie, genetic subtype classification in single-domain generalization setting, and tumor identification in multi-domain generalization setting) using digitized histopathology images in computational pathology.

We speculate that STRAP helps models learn domain-agnostic and class-specific visual representations by removing the original texture and/or high-frequency components from the histopathology images, which are domain-specific and class-irrelevant, and predominantly leaving shape-biased and/or low-frequency content, which are domain-irrelevant and class-specific. In fact, more intensive style transfer with a higher stylization coefficient resulted in superior performance. Furthermore, when tested on stylized version of the out-of-distribution test dataset with random test-time styles, STRAP with Artistic Paintings showed significantly higher performance compared to the baseline SA and SN approaches. Also, our experiments on the low-frequency components demonstrated that the STRAP approach helps models exploit lower frequency components, in contrast to the standard SA and SN approaches that rely more on higher frequency components. This speculation is also consistent with the hypotheses proposed by Geirhos \etal~\cite{Geirhos2018-xg} and Wang \etal~\cite{Wang2020-jq}—that shape-biased and/or low-frequency features are essential for deep learning models to learn robust and generalizable visual representations.

To the best of our knowledge, no previous study has applied medically-irrelevant image manipulation for the development of deep learning models for medical imaging. Four previous studies have applied the style transfer technique to medical imaging tasks in computational pathology~\cite{Cicalese2020-rs, Shin2021-vg} and skin lesion classification~\cite{Mikolajczyk2019-he, Nyiri2020-oz}. However, these studies employed medically-relevant transformation with the aim of combating data scarcity, class imbalance, and stain variation. Our study demonstrates that medically-irrelevant transformation, \ie, STRAP with Artistic Paintings or Natural Imaging style sources, can result in improved performance and generalizability, when compared with medically-relevant transformation, \ie, style transfer with Histopathologic Imaging style source and stain augmentation. A possible explanation for this phenomenon is that medically-irrelevant style transfer can result in a wider variety of transformation using a more diverse set of styles compared to the medically-relevant approaches for which the variations in color and texture are more uniform and thus, limited. Tobin \etal~\cite{Tobin2017-hi} showed that an object detection model that generalizes to real-world images can be trained by using unrealistic simulated images with a diverse set of random textures, rather than by making the simulated images as realistic as possible. As in the human learning process, learning class-specific and domain-irrelevant patterns from data is essential for deep learning models, and the style transfer technique with a diverse set of random styles can be a powerful tool to control the representations models learn. 

Although data augmentation is widely used when training deep learning models for medical imaging tasks, its potential has not yet been fully studied and still remains an active area of research. Moreover, an optimal configuration of data augmentation methods may vary among applications. As our study suggests, data augmentation can be a simple yet powerful tool for learning domain-agnostic representation. Further research is warranted to identify optimal data augmentation techniques for a variety of medical imaging tasks, and medically-irrelevant transformations such as the proposed STRAP approach should be considered, along with established methods.

As shown in Table \ref{tab:r_msi}, STRAP with Artistic Paintings achieved higher performance in the out-of-distribution setting, compared to the in-distribution setting, whereas opposite results were observed for the other baseline approaches and state-of-the-arts. As described in Section \ref{subsubsec:taskone}, the training data in the in-distribution setting was a multi-source domain dataset, whereas the training data used for the out-of-distribution setting was a single-source dataset. Although it is often said that diverse multi-institutional datasets are needed for training models that generalize on unseen data~\cite{Kelly2019-xn}, our study may suggest that a well-curated homogeneous dataset could provide value in training domain-agnostic models, if a model has sufficient capability to learn domain-invariant and class-specific representations, similar to the way in which humans learn from a set of representative examples (\eg, content presented in textbooks).

Besides supervised learning, our approach may be applicable to self-supervised learning. A contrastive learning framework, such as SimCLR~\cite{Chen2020-vr} and MoCo~\cite{He2020-vw}, learns representations by maximizing agreement between differently augmented views of the same data example via a contrastive loss (thus, relying heavily on a stochastic data augmentation module). Chen \etal~\cite{Chen2020-vr} showed that the composition of data augmentation operations is crucial in yielding effective representations, and that unsupervised contrastive learning benefits from strong data augmentation. In medical imaging, contrastive learning may require a tailored composition of data augmentation operations, and our medically-irrelevant STRAP has the potential to serve as one of the core transformation operations.

One limitation of this study is that we only tested our approach with classification tasks in the field of computational pathology. Further studies are warranted to investigate whether our approach could prove its efficacy and robustness 1) for non-classification tasks such as detection, segmentation, and survival prediction, and 2) in other medical imaging domains, such as radiology, ophthalmology, and dermatology. Another limitation is STRAP's relatively longer runtime compared to the other two baseline approaches, where the average runtime was 1.08 s for STRAP, 8.13 ms for SA, and 6.42 ms for SN on a workstation with a GeForce RTX 2080 Ti (NVIDIA, Santa Clara, CA) graphics processing unit, a Core i9-9820X (10 cores, 3·3 GHz) central processing unit (Intel, Santa Clara, CA, and 128 GB of random-access memory. Improvement in computational efficiency is required to apply STRAP as one of on-the-fly data augmentations.

In conclusion, we have introduced STRAP, a form of data augmentation based on random style transfer with medically-irrelevant style source, for learning domain-agnostic visual representations in computational pathology. Our experiments demonstrated that our approach yields significant improvements in test performance on classification tasks in computational pathology, particularly in the presence of domain shift. Our study provides evidence that 1) CNNs are reliant on low-level texture content and are therefore vulnerable to domain shifts in computational pathology, and that 2) medically-irrelevant STRAP can be a practical tool for mitigating that reliance and, therefore, a possible solution for learning domain-agnostic representations.

\section*{Acknowledgements}

This work was funded by the Stanford Departments of Biomedical Data Science and Pathology, through a Stanford Clinical Data Science Fellowship to RY. We would like to thank Blaine Burton Rister for detailed and valuable feedback on the manuscript. We would also like to thank Nandita Bhaskhar, Khaled Kamal Saab, and Jared Dunnmon for their helpful discussions.

\bibliographystyle{unsrt}  
\bibliography{main}

\end{document}